\newdimen\figwidth
\documentclass{elsart}
\usepackage{amssymb}
\usepackage{epsfig}
\begin{document}
\begin{frontmatter}
\title{Solvable  Chaos}
\author[gmpib]{B. Grammaticos}
\ead{grammati@paris7.jussieu.fr}
\author[poly]{A. Ramani}
\ead{ramani@cpht.polytechnique.fr}
\author[lpthe]{C.M. Viallet}
\ead{viallet@lpthe.jussieu.fr}
\address[gmpib]{GMPIB, Universit\'e Paris VII,  Tour 24-14, 5$^e$\'etage,
case 7
021,\\ F--75251 Paris Cedex 05}
\address[poly]{Centre de Physique Th\'eorique, UMR 7644 
\\ \relax Ecole Polytechnique, 91128 Palaiseau, France}
\address[lpthe]{Laboratoire de Physique Th\'eorique et des Hautes Energies,
\\ \relax 4 Place Jussieu, Boite 126, F--75252 Paris Cedex 05}

\begin{abstract}
We present classes  of discrete reversible systems which are at the
 same time  chaotic  and solvable.
\end{abstract}
\end{frontmatter}
\maketitle

\section{Introduction}

Chaos and solvability are antithetical notions and their coexistence
in a dynamical system may sound paradoxical. The term chaos is
traditionally used to designate systems which exhibit crucial
dependence on the initial conditions manifested through exponentially
diverging trajectories and extreme instability~\cite{EcRu85}. On the
other hand, solvability is often associated to smooth, regular
behaviour, related to the existence of invariants, and is usually
coming from integrability.

In what follows we shall show that the explicit solvability of a
mapping is not incompatible with a chaotic behaviour. This is done in
the same spirit as in~\cite{Um97,Um97b}, but we will give example of
reversible systems (i.e. there exists a similarity transformation
between the forward evolution and the backward evolution), exhibiting
at the same time features of chaos (e.g. sensitive dependence on the
initial conditions, positive algebraic entropy, ergodicity), and which
are solvable. Our point is neither to recall the possible coexistence
of KAM tori and chaotic regions which is well known~\cite{RoQu92}, nor
to give the ultimate definition of integrability, solvability or
chaos.

Our point is to present reversible systems which lie on the border of
solvability/integrability and chaos.

\section{Specific  examples}

We will use maps which are constructed from recurrences, that is to
say sequences where each term is given as a function of the previous ones:
\begin{eqnarray}
\label{recurrence}
 x_{n+1} = f ( x_n, x_{n-1}, \dots, x_{n-\nu} )
\end{eqnarray}
If $x_{n+1}$ is a function of only $x_n$, we have a one dimensional
map, also called a system of order one. If $\nu = 1$ (resp. $\nu=r$)
we may define from (\ref{recurrence}) a map in two dimensions
(resp. $r+1$ dimensions) by
\begin{equation}
[ x_n, x_{n-1}, \dots, x_{n-\nu} ] \longrightarrow [ x_{n+1}, x_{n},
\dots, x_{n - \nu +1} ]
\end{equation}
The space of initial conditions is of dimension $\nu + 1$.

\subsection{One dimensional maps}

We start with a well-known one dimensional example to illustrate
the fact that solvability is compatible with chaotic behaviour:
\begin{eqnarray}
x_{n+1}=2\; x_n^2-1 \label{logistic}
\end{eqnarray}
This map is know to be chaotic~\cite{CoEc80}.  Using the similarity
between (\ref{logistic}) and the doubling rule for cosine: $\cos
(\,2\; \omega) =2\cos^2 \omega-1$, the solution of (\ref{logistic}) is
given by $x_n=\cos 2^n\alpha$ where $\alpha$ is some constant
determined by the initial conditions. We shall not dwell upon the
exponentially fast loss of the memory of the initial conditions due to
the presence of the $2^n$ factor: the existing literature covers the
topic in an exhaustive way~\cite{EcRu85}. It remains that we have here
an example of a system that is explicitly solvable and which has a
chaotic behaviour.

The previous example is not an isolated occurrence. Whole families of
first-order (one-dimensional) systems exist which are solvable while
at the same time exhibiting exponentially fast loss of the memory of
the initial conditions.  There exist a number of results in this
direction for maps of the interval~\cite{Ve91,Um97,Um97b}.  It was
shown that the only nonlinear polynomial maps where the solution can
be explicitly given are the map $x_{n+1}=x_n^2$ and (\ref{logistic}),
up to a homographic transformation.

For rational maps his classification is based on results of
Ritt~\cite{Ri23}.  For degree one, the only solution is the
homographic map. For degree 2, there exist 8 different recurrences,
the simplest ones being
\begin{eqnarray}
x_{n+1} & = &{1\over 2i}(x_n-{1\over x_n}) \label{3} \\
x_{n+1} & = & -{1\over 4}(x_n-2+{1\over x_n}) \label{4}
\end{eqnarray}
Their solutions can be given in terms of the Weierstra{\ss} elliptic
function $\wp(z)$, defined through the equation
$(\wp')^2=4\wp^3-4\wp$.  Equation (\ref{3}) has solution
$x_n=\wp((1+i)^nz)$.  Similarly the solution of (\ref{4}) is
$x_n=\wp^2((1+i)^nz)$, and so on.

The major problem with the above mappings is that they are not
invertible. While one can define the image of a given point in an
unambiguous way, the same is not true for the inverse
evolution. Moreover the number of preimages of a point grows
exponentially fast (a property which was deemed incompatible with
integrability~\cite{GrRaTa94}). In the next paragraph we will exhibit
{\em reversible} maps, defined by {birational} transformations showing
 features of chaos and which are solvable.  One
condition to construct such maps is to consider transformations in
more than one dimension.

\subsection{Two-dimensional maps}

In the case of transformation (\ref{logistic}), the key ingredient was
the doubling relation $\omega_{n+1}=2\omega_n$ (or a higher multiple
$\omega_{n+1}=k\omega_n$). We start from a linear equation:
\begin{eqnarray}
\omega_{n+1}+\omega_{n-1}=k\omega_n
\label{7}
\end{eqnarray}
The solution of (\ref{7}) is straightforward:
$\omega_n=a\lambda_+^n+b\lambda_-^n$ where
$\lambda_\pm=(k\pm\sqrt{k^2-4})/2$. Exponentiating (\ref{7}) and
setting $x=e^\omega$ we obtain the recurrence:
\begin{equation}
\label{8}
x_{n+1}x_{n-1}=x_n^k
\end{equation}
 The chaotic character of this map (for $k>2$) can be assessed easily
through the computation of its algebraic entropy~\cite{BeVi99}. This
quantity is a measure of the complexity of the map and is given by
$\epsilon=\lim_{n\to \infty}
\log(d_n)/n$ where $d_n$ is the degree of the $n$-th iterate. In the
case at hand we find that $\epsilon=\log(\rho_1)$, with the $\rho_1$
the larger of the two roots defined above. This leads to $\epsilon>0$
for $k>2$. In fact the same value for the entropy is obtained for all
the maps derived from (\ref{7}).

While map~(\ref{8}) is rather trivial, it is possible to construct
a much more interesting one by setting 
\begin{equation}
\label{transf}
x_n=\tan\omega_n
\end{equation}
  We get the recurrence:
\begin{equation}
\label{9}
{x_{n+1}+x_{n-1}\over 1-x_{n+1}x_{n-1}}=f_k(x_n)
\end{equation}
 where $f_k$ is a rational function of $x_n$ depending on the value of
$k$. It is simply the expression of $\tan k\omega_n$ in terms of
$\tan\omega_n\equiv x_n$. For the first few values of $k$ we have
$f_1(x_n)=x_n$, $f_2(x_n)=2x_n/(1-x_n^2)$,
$f_3(x_n)=(3x_n-x_n^3)/(1-3x_n^2)$ etc.  The case $k=1$ corresponds to
a trivial map which is periodic with period 3. The case $k=2$ can
be easily integrated. There exists an invariant
$c=(x_{n-1}-x_{n})/(1+x_{n-1}x_{n})$, which allows to reduce the
map to a homographic one and solve it completely.

The case $k=3$ is more interesting. Indeed the map constructed
from the recurrence
\begin{eqnarray}
x_{n+1}={3x_n-x_n^3-x_{n-1}(1-3x_n^2)\over
1-3x_n^2+(3x_n-x_n^3)\, x_{n-1}}
\label{tg}
\end{eqnarray}
is both chaotic and solvable (since it is linearizable).  The
 generating function of the sequence of degrees~\cite{FaVi93} may
 easily be inferred from the first terms of this sequence:
\begin{eqnarray}
g={1+2s+s^2-2s^3+s^4-2s^5\over(1-s)(1+s+s^2)(1-3s+s^2)}
\label{11}
\end{eqnarray}
The value of the algebraic entropy $\epsilon = \log ((3+\sqrt{5})/2)$
is read off from~(\ref{11}).  This positive entropy is a sign of
chaotic behaviour.  

Similar maps can be constructed for $k \ge 3$.  They will have a positive
algebraic entropy and be linearizable.

\section{Graphical analysis}

Figure 1 shows the orbit of an arbitrary initial point under $10^6$
iterations of the evolution~(\ref{tg}), plotting the pairs $(x_n,
x_{n+1})$ in the two-dimensional plane.  It shows that the orbits tend
to fill phase space.

\par\smallskip
%\par\smallskip
\par\bigskip
\epsfxsize=\figwidth
\centerline {\epsffile[0 0 512 512]{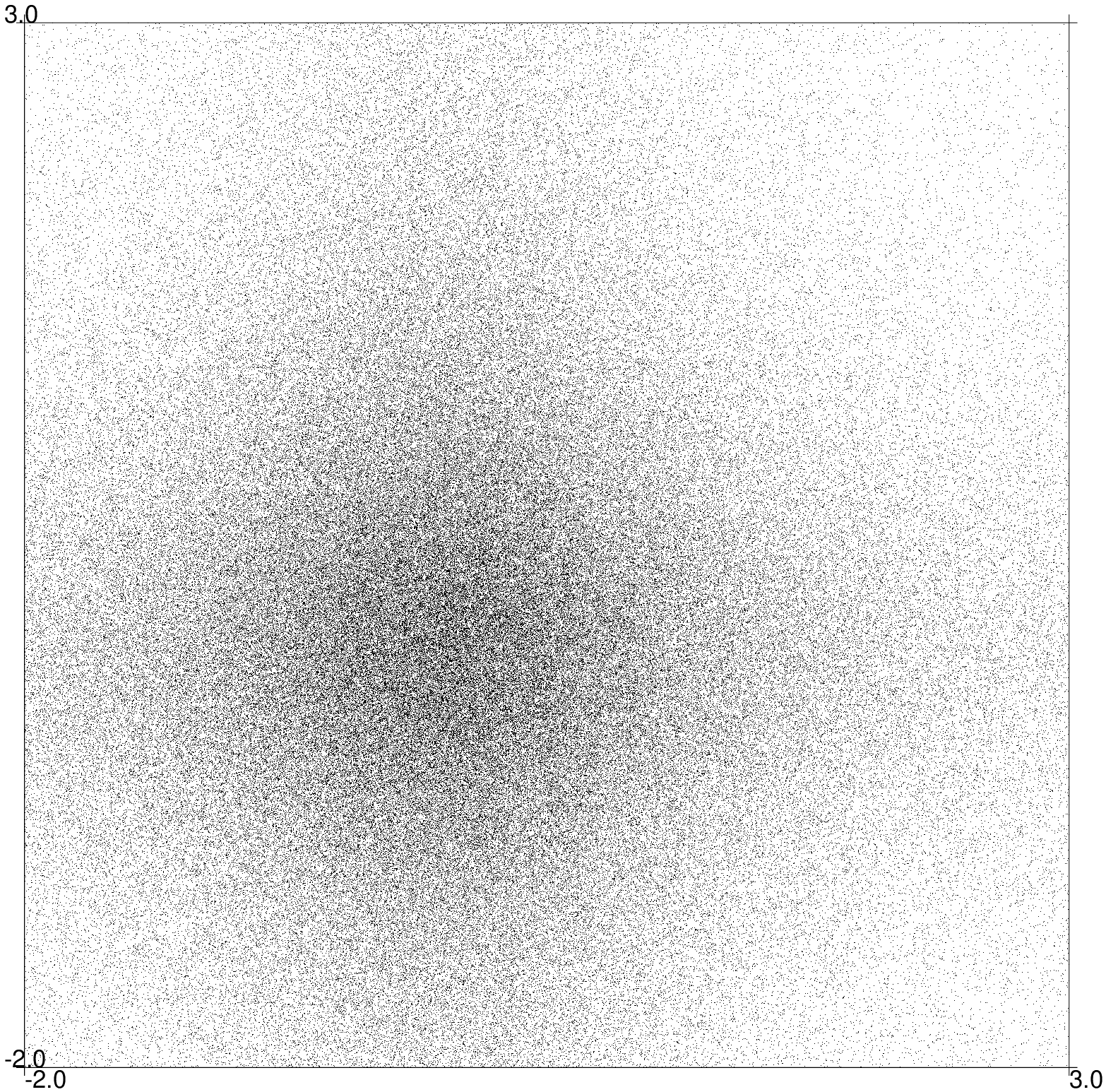}}
\centerline{\it  Figure 1: A typical orbit of a point under map (\ref{tg})}
\par\smallskip

Figure 2 shows the images of a segment under a dozen of iterations
of~(\ref{tg}).  Pushing the iteration further does not change the
qualitative features of the image. It just increases the density of
lines.  

\par\bigskip
\epsfxsize=\figwidth
\centerline {\epsffile[0 0 512 512]{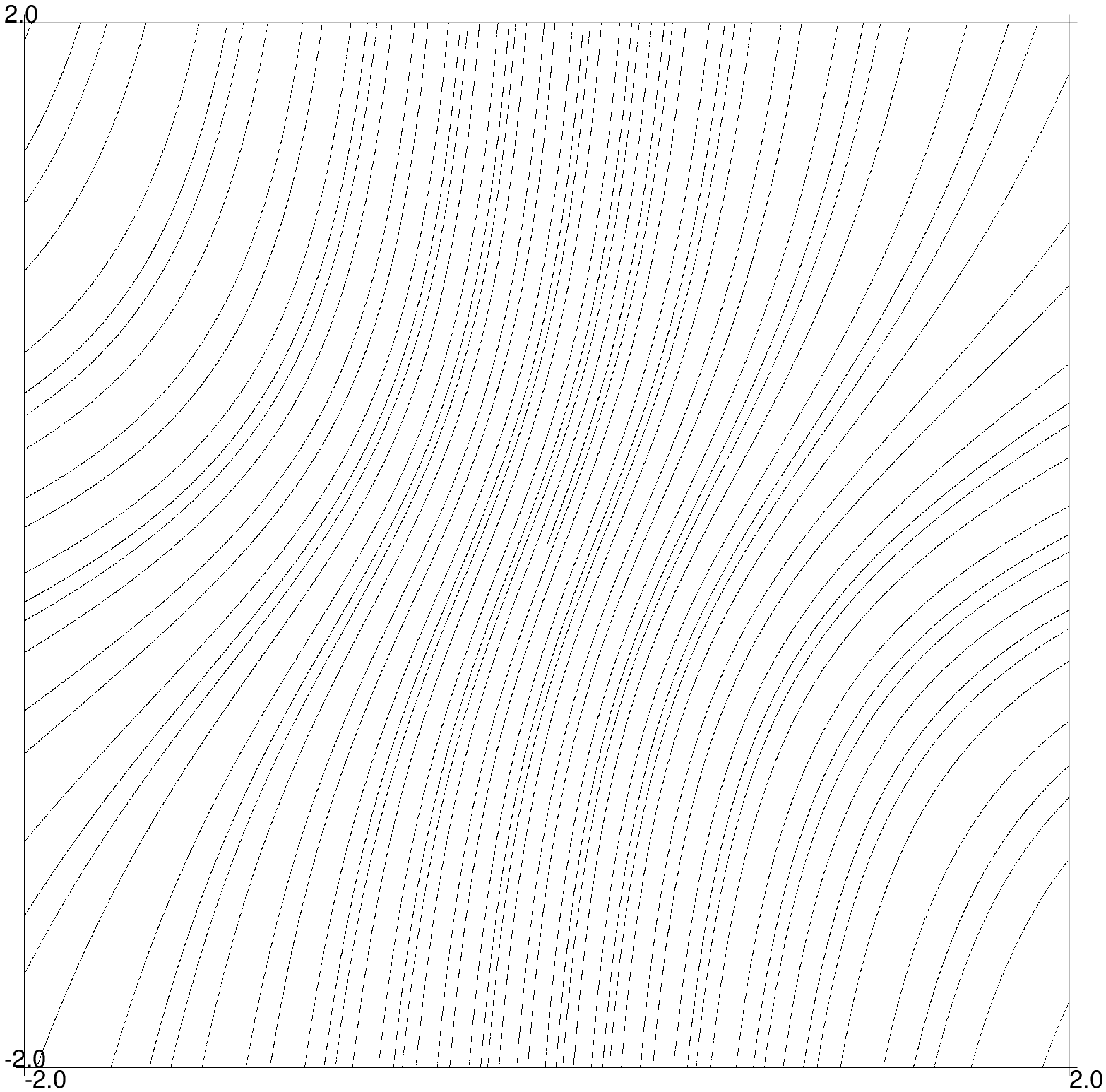}}
\centerline{\it  Figure 2: Images of a segment under map (\ref{tg})}
%\par\bigskip
\par\smallskip

The fact that the map (\ref{tg}) has an algebraic entropy of
$\log((3+\sqrt{5})/2)$ invites us to compare it to the map described
in~\cite{HiVi98}, which has the same algebraic entropy, and is
chaotic:
\begin{eqnarray}
x_{n+1}+x_{n-1}=x_n+{a\over x_n^2}\label{hv}
\end{eqnarray}
with $a$ a constant.
\par\bigskip
\par\smallskip
\epsfxsize=\figwidth
\centerline {\epsffile[0 0 512 512]{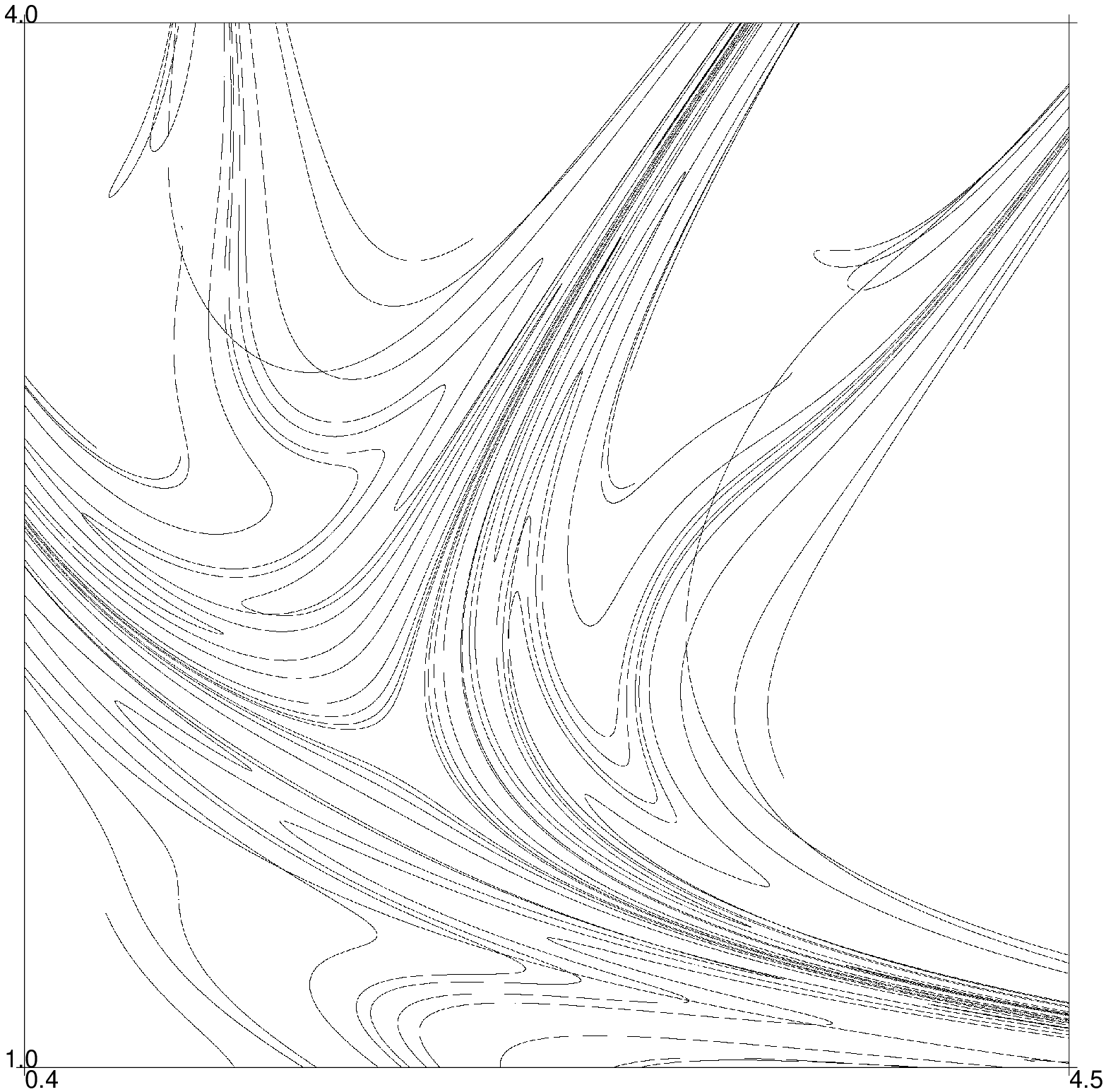}}
\centerline{\it  Fig. 3: Images of a segment under map (\ref{hv})}
\par\bigskip

The typical orbit of an arbitrary point was shown in~\cite{HiVi98},
presenting large chaotic regions, and we do not reproduce it here.

Figure 3 shows  the images of a segment of a straight line under
$27$  iterations of~(\ref{hv}), and is to be compared with Figure 2.
The overall pictures differ qualitatively.  A detailed analysis of the
behaviour of the iterates shows that in the case of the map
(\ref{hv}) the evolution is slow till the points get near one of the
singularities, in which case they get a major boost, which results in
the rich structure of Figure 3. On the contrary, in the case of
map~(\ref{tg}) neighboring points are uniformly repelled and the
overall results are regular and smooth.  Both have a strong dependence
on the initial conditions, but in different ways.

All numerical calculations leading to Figures 2 and 3 have been
performed with multi-precision~\cite{gmp} arithmetics so as to
guarantee the reliability of the results, and a test was performed on
all image points: after an evolution of $N$ steps, we ``go back in
time'' the same number of steps to recover the initial data. We
adjusted the precision in such a way that the true initial data and
the result of this round trip did not differ by more than $10^{-3}$.

The distinguishing feature of (\ref{tg}) compared to (\ref{hv}) if
that (\ref{transf}) has an integral
\begin{equation}
\label{conique}
(\omega_{n+1}- \lambda_+ \; \omega_{n}) (\omega_{n+1}- \lambda_- \;
\omega_{n}) = cst
\end{equation}
with $\lambda_\pm = ( 3 \pm \sqrt{5} )/2$.  The $\omega$-plane is
foliated by invariant curves, but  the picture is scrambled by the
transformation (\ref{transf}) since the tangent function is periodic.
It is easy to understand the aspect of Figure~1. The
curve~(\ref{conique}) is transformed by~(\ref{transf}) into
\begin{equation}
(\arctan {x_{n_+1}} - \lambda_+ \; \arctan{x_n} ) (\arctan{x_{n+1}} -
\lambda_- \;\arctan{x_n} ) = \kappa
\label{newconique}
\end{equation}
with $x_{n+1}$ and $x_{n}$ the two coordinates of the plane of
Figure~1, and $\kappa$ a constant. Such a curve cuts a line $x_n=\xi$
at an infinite dense set of points.

One may notice that there exist two holomorphic foliations which are
left invariant by (\ref{9}), as in the analysis of~\cite{CaFa02}, and
our construction exemplifies their result.

One interesting feature of transformation~(\ref{transf}) is that it is
 not a mere change of coordinates. It takes a birational map into a
 birational one, changing the algebraic entropy.

\section{Arithmetical analysis}

The difference between maps (\ref{tg}) and (\ref{hv}) can be
illustrated by an analysis based on the approach recently introduced
by Roberts and Vivaldi~\cite{RoVi03}. These authors have studied the
effect of the existence of rational integrals of motion for rational
maps when the evolution is considered over a finite field. The
simplest realization of such an evolution is through integer
arithmetics modulo some prime integer $p$.  The basic observation
of~\cite{RoVi03} (see also~\cite{RoJoVi03}) is that if there exists a
rational invariant, the orbit is confined to an algebraic curve, and
the genus $g$ of this curve is at most $g\leq 1$, if the original map
is of infinite order.  Such curves over finite fields have a maximum
number of points (the Hasse-Weil bound $HW(p,g) = p+1+2\; g \;
\sqrt{p} $), and as a consequence the number of points on the same
orbit is also bounded by $HW(p,g)$. In short: algebraically integrable
maps have a large number of orbits and those are short. Chaotic maps
have a smaller number of orbits and they are longer. Notice that this
fits with the idea that chaotic orbits may explore the whole phase
space, contrary to what happens in the integrable case.

We have performed a sampling of initial points for increasing values
of $p$ and plotted the mean value of the length of orbits (with the
rule to terminate the iteration when meeting a singular point or
closing a loop), for the maps~(\ref{tg}), (\ref{hv}), and one
additional map which is known to be algebraically integrable (so
called McMillan map), given by the recurrence:
\begin{eqnarray}
x_{n+1} +  x_{n-1} = 2 \; a {{x_n}\over{ x_n^2-1}}
\label{mcm}
\end{eqnarray}
with $a$ a free parameter.  

\figwidth= 12 truecm

\par\bigskip
\epsfxsize=\figwidth
\centerline {\epsffile[0 0 512 512]{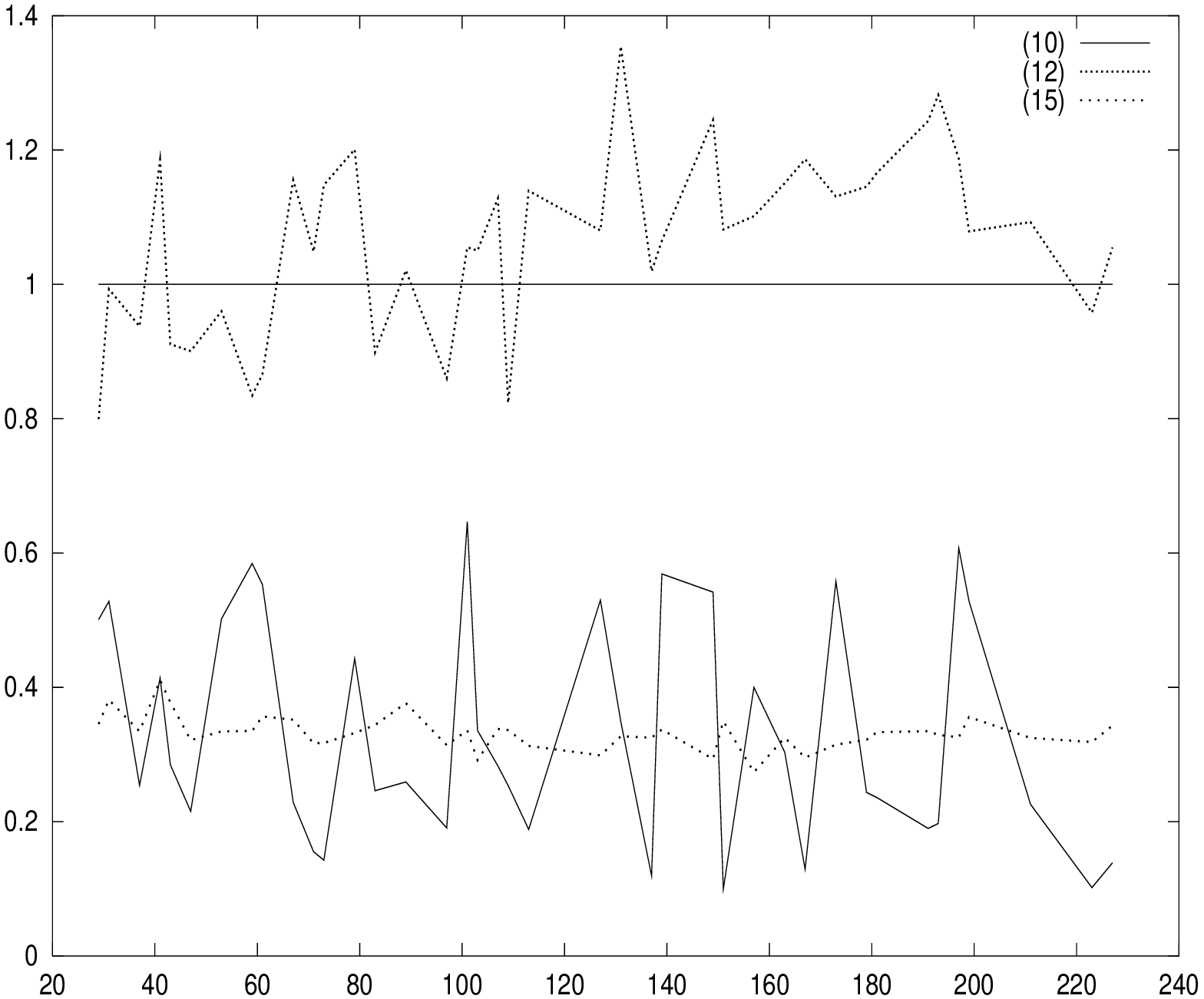}}
\centerline{\it  Figure 4: Mean length of orbits vs $p$}
\par\bigskip

Figure 4 shows the values of the mean length for increasing $p$,
normalized by dividing by $HW(p,1) = p+1+2\; \sqrt{p}$, for maps
(\ref{tg}), (\ref{hv}), and (\ref{mcm}).  It discriminates between
(\ref{tg}) and (\ref{hv}) by showing that the former verifies a
virtual Hasse-Weil bound while the latter does not. The curve
corresponding to map~(\ref{mcm}) is there for reference, as well as
the line corresponding to a constant normalized length of
$1$. Clearly, although the orbits are not confined to any invariant
algebraic curve, the arithmetic test of~\cite{RoVi03} places the
map~(\ref{tg}) in the integrable class together with~(\ref{mcm}), in
contrast to~(\ref{hv}).

A more detailed analysis of the statistics of the length of orbits
will be done elsewhere.

\section{Generalizations}

We can construct generalizations of~(\ref{9}), based on the properties
of elliptic functions. Introduce $x_n=\wp(\omega_n)$ and
$y_n=\wp'(\omega_n)$ where $\wp$ is the Weierstra{\ss} elliptic
function with elliptic invariants $g_2$ and $g_3$. The addition
formulae for the Weierstra{\ss} elliptic function yield:
\begin{eqnarray}
 \nonumber
x_{n+1} & = &{1\over 4}\left({h_k(x_n,y_n)+y_{n-1}\over
f_k(x_n,y_n)-x_{n-1}}\right)^2-f_k(x_n,y_n)-x_{n-1} \\ y_{n+1} & = &
{x_{n+1}(h_k(x_n,y_n)+y_{n-1})}\over{ x_{n-1}-f_k(x_n,y_n)}
\label {14} \\ \nonumber
&& \quad - \quad {{f_k(x_n,y_n)y_{n-1} + h_k(x_n,y_n)x_{n-1}}\over{
x_{n-1}-f_k(x_n,y_n)}}
\end{eqnarray}
 where the functions $f_k$ and $h_k$ are the expressions of
$\wp(k\omega_n)$ and $\wp'(k\omega_n)$ in terms of $x_n=\wp(\omega_n)$
and $y_n=\wp'(\omega_n)$.  

We have to ensure that both $(x_n,y_n)$ and
$x_{n-1}, y_{n-1}$ lie on the same elliptic curve i.e.
\begin{eqnarray}
\label{15a}
y_n^2 & = & 4x_n^3-g_2x_n-g_3 \\
y_{n-1}^2 & = & 4x_{n-1}^3-g_2x_{n-1}-g_3
\label{15b}
\end{eqnarray}

Equations (\ref{15a},\ref{15b}) together with the formula for the
duplication, triplication,~... of the arguments yield the value of the
$f_k$ and $h_k$. Iteration (\ref{14}) defines a map in four
variables which has two algebraic invariants given by solving
(\ref{15a},\ref{15b}) in terms of $g_2,g_3$:

\begin{eqnarray}
\label{16a}
g_2 & = & {y_{n-1}^2-y_n^2\over
x_n-x_{n-1}}+4(x_{n-1}^2+x_nx_{n-1}+x_n^2) \\
g_3 & = & {x_{n-1}y_{n}^2-x_ny_{n-1}^2\over
x_n-x_{n-1}}-4x_nx_{n-1}(x_{n-1}+x_n)
\label{16b}
\end{eqnarray}

For $k=2$ we have $f_2=-2x+z^2/(4y^2)$ and $h_2=-y+3xz/y-z^3/(4y^3)$
where $z$ stands for $\wp''(\omega_n)=6\wp^2(\omega_n)-g_2/2\equiv
6x^2-g_2/2$. In this case there exists an additional invariant
\begin{eqnarray}
C=\left({y_{n-1}+y_n\over
x_n-x_{n-1}}\right)^2-4(x_n+x_{n-1})
\label{17}
\end{eqnarray}
and the map is integrable with vanishing entropy (quadratic growth
of the degree).

For $k=3$ we have $f_3=x+4y^2(12xy^2z-4y^4-z^3)/(12xy^2-z^2)^2$ and
$h_3=-y-4y(12xy^2z-8y^4-z^3)(12xy^2z-4y^4-z^3)/(12xy^2-z^2)^3$.  This
case, and actually all cases with $k \geq 3$, have positive entropy
and solvability is ensured through the relation to (\ref{7}).

\section{Conclusion}

We have shown that there exist infinite families of rational maps
which, at the same time, have positive algebraic entropy,  present
features of chaos, and are solvable.  Their solvability is related to
a reduction to a linear equation through the appropriate non rational
transformations, but they remain reversible.

While the examples we exhibited here are based on specific Ans\"atze,
there exist infinite families of solvable mappings with positive
algebraic entropy.  As a matter of fact one could perform the same
derivation using any function for which one can express $f(x+y)$ in
terms of $f(x)$ and $f(y)$.

Open questions remain, like what is the meaning of the statistics of
the length of orbits in the arithmetic approach
of~\cite{RoVi03,RoJoVi03}. We will return to that in some future
publication.

Acknowledgments:
We acknowledge stimulating discussions with C. Favre.

%\bibliography{/users/lpthe/viallet/papiers/ref}

\end{document}